\documentclass[superscriptaddress,nofootinbib,10pt,preprintnumbers,notitlepage]{revtex4-1}
\usepackage{epsfig,amsmath,amssymb}
\usepackage{color}
\usepackage{float}
\usepackage[colorlinks]{hyperref}
\usepackage{subfigure}
\usepackage[utf8]{inputenc}

\hypersetup{
    linktocpage,
     colorlinks,
     citecolor=darkgreen,
     linkcolor= darkgreen,
     urlcolor=darkgreen
}
\definecolor{darkred}{rgb}{0.5,0.0,0.0}
\definecolor{darkblue}{rgb}{0.0,0.0,0.9}
\definecolor{darkerblue}{rgb}{0.0,0.0,0.5}
\definecolor{purple}{rgb}{0.5,0.0,0.5}
\definecolor{darkgreen}{rgb}{0.0,0.5,0.0}
\definecolor{black}{rgb}{0.0,0.0,0.0}
\definecolor{brown}{rgb}{0.6,0.4,0.2}
\definecolor{newpurple}{rgb}{0.65, 0.38, 0.61}
\definecolor{newyellow}{rgb}{0.9718, 0.6093, 0.0759}
\definecolor{amber}{rgb}{1.0, 0.75, 0.0}
\definecolor{newblue}{rgb}{0.4, 0.52, 0.85}
\definecolor{newred}{rgb}{0.8524, 0.2595, 0.3294}
\definecolor{newgreen}{rgb}{0.2, 0.8, 0.2}
\definecolor{SMgreen}{rgb}{0.56, 0.69, 0.19}
\definecolor{neworange}{rgb}{0.94, 0.462, 0.162}

\definecolor{BrickRed}{rgb}{0.9,0.1,0}

\def\Re{\mathcal{R}e}

\newcommand{\bea}{\begin{eqnarray}}
\newcommand{\eea}{\end{eqnarray}}
\newcommand{\beq}{\begin{equation}}
\newcommand{\eeq}{\end{equation}}
\newcommand{\ec}{\end{center}}
\newcommand{\bc}{\begin{center}}

\begin{document}

\preprint{DO-TH 16/32}
\title{Revisiting $B\to K^\ast( \to K\pi) \nu\bar{\nu}$ decays}
\author{Diganta Das}
\email[Electronic address:]{diganta@prl.res.in}
\affiliation{Physical Research Laboratory, Navrangpura, Ahmedabad 380 009, India}
\author{Gudrun Hiller}
\email[Electronic address:]{ghiller@physik.uni-dortmund.de}
\author{Ivan Ni\v sand\v zi\' c}
\email[Electronic address:]{ivan.nisandzic@tu-dortmund.de}
\affiliation{Institut f\" ur Physik, Technische Universit\" at Dortmund, D-44221
Dortmund, Germany}

\begin{abstract}
The rare decay  $B\to K^\ast( \to K\pi) \nu\bar{\nu}$ is expected to play an
important role in searches for physics beyond the Standard Model  at the near future
$B$-physics experiments. We investigate  resonant and non-resonant backgrounds  that
arise beyond the narrow-width approximation for the $K^\ast$.  
Non-resonant $B\to  K\pi  \nu\bar{\nu}$ decays are analyzed  in the  region of low
hadronic recoil, where  $B \to K \pi$ form factors from
Heavy-Hadron-Chiral-Perturbation Theory are available. In a Breit-Wigner-type model
interference-induced  effects in the  $K^*$ signal region are found to be sizable, 
as large as $20\%$  in the branching ratio.
Corresponding effects in  the longitudinal polarization fraction $F_L$ are smaller,
at most around few \%. Effects of the broad scalar states $K_0^\ast$ and $\kappa$
are at the level of percent in the branching fraction in the $K^\ast$ signal region
and negligible in $F_L$. 
Since the backgrounds to $F_L$ are  small this observable constitutes a useful probe
of form factors calculations, or alternatively, of right-handed currents in the
entire $q^2$-region.
The forward-backward asymmetry in the $K \pi$-system, $A_{\rm FB \, L}^K$, with
normalization to the longitudinal decay rate probes predominantly S,P-wave
interference free of short-distance coefficients and
can therefore be used to control the resonant and non-resonant  backgrounds.
\end{abstract}
\maketitle

\section{Introduction}

The rare semi-leptonic hadron decays induced by $|\Delta B|=| \Delta S|=1$ flavor changing neutral currents are sensitive probes of the Standard Model (SM) and beyond.
The transitions $b\to s\ell^+\ell^-$, where $\ell=e,\mu$,   have been the subject  of extensive theoretical and experimental studies in the past several decades~\cite{Blake:2015tda, Blake:2016olu}. 
The main theoretical challenges for the reliable extraction of Wilson coefficients from the experimental data arise from the requirement of the quantitative understanding of QCD backgrounds at large distances. 
To test the SM  and  to improve the understanding of theoretical uncertainties  one can pursue  studies with $b\to s\nu\bar{\nu}$ transitions, which are related by $SU(2)_L$ to $b \to s \ell^+ \ell^-$ but not being subjected to sizable electromagnetic contributions from charm quarks.
While  dineutrino modes are theoretically better understood, they are  experimentally more challenging, and have not been observed to date. One can expect, however, that exclusive  dineutrino modes with SM branching ratios of  $\sim 10^{-5}$ will be observed and probed  at the forthcoming Belle II experiment~\cite{Aushev:2010bq, BelleII}.
The current best  limit is from the Belle collaboration  and reads, at 90 \% confidence level,~\cite{Grygier:2017tzo}
\begin{align}
\mathcal{B}(B\to K^{\ast 0}\bar{\nu}\nu)  < 1.8 \times 10^{-5} \,  ,
\end{align}
which is just around the corner of the SM prediction.
Dedicated studies of the impact of  new physics on $b\to s \nu\bar{\nu}$ processes can be found in the recent literature~\cite{Colangelo:1996ay, Buchalla:2000sk,Altmannshofer:2009ma, 
Bartsch:2009qp, Buras:2014fpa}, see also~\cite{Girrbach-Noe:2014kea, Niehoff:2015qda}. Here we focus on $B^0\to K^{0\ast}( \to K \pi )\nu\bar{\nu}$ decays and analyze the interplay of the SM induced backgrounds for a $K^*$-meson beyond the narrow-width approximation (NWA). Corresponding effects in $B \to K^* ( \to K \pi ) \ell^+ \ell^-$ decays 
from  scalar states and non-resonant contributions have been investigated previously in~\cite{Becirevic:2012dp,Matias:2012qz, Blake:2012mb} and~\cite{Das:2014sra,Das:2015pna}, respectively. Interestingly, the S-wave fraction in $B \to K^* ( \to K \pi ) \mu^+ \mu^-$ has recently been measured by the LHCb collaboration~\cite{Aaij:2016flj}.

We consider only decays of neutral $B$-mesons and omit the charge indices throughout; the corresponding decays of  charged $B$-mesons are additionally impacted by tree level  charged currents via a resonant tau lepton~\cite{Kamenik:2009kc}.

After setting the notation in Sec.~\ref{Effective Hamiltonian}, we give amplitudes and distributions for  an asymptotic final state $K^\ast$ in Sec.~\ref{Onshell}. In Sec.~\ref{scalars} we work out  effects from intermediate scalar mesons $K_0^{\ast}$ and $ \kappa$, which contribute to the creation of the outgoing $K\pi$ pair beyond the NWA for $K^\ast$. 
Non-resonant $B \to K \pi \nu\bar{\nu}$ contributions for a $K\pi$-mass around the one of the $K^\ast$ are analyzed  in Sec.~\ref{Non-resonant contributions}.
In Sec.~\ref{sec:con} we conclude. Auxiliary information is deferred to three appendices.

\section{Generalities}

We give the effective $b\to s\nu\bar{\nu}$ Hamiltonian  used in this work and some notation in Sec.~\ref{Effective Hamiltonian} and  $B\to K^\ast\nu\bar{\nu}$ distributions for
a zero-width $K^*$ in Sec.~\ref{Onshell}.

\subsection{Effective Hamiltonian and notation}\label{Effective Hamiltonian}
We begin with the low energy effective Hamiltonian for  $b\to s\nu\bar{\nu}$ transitions following~\cite{Colangelo:1996ay, Altmannshofer:2009ma} 
\begin{equation}
\mathcal{H}_{\text{eff}}=-\frac{4G_F}{\sqrt{2}}\lambda_t\frac{\alpha}{8\pi}\bigg[(C_L+C_R)(\bar{s}\gamma_\mu b)+(C_R-C_L)(\bar{s}\gamma_\mu\gamma_5 b)\bigg]\sum_i \bar{\nu}_i\gamma^\mu(1-\gamma_5)\nu_i+\text{h.c.},
\end{equation}
where $\lambda_t=V_{tb}V_{ts}^\ast$  is  the product of the Cabbibo-Kobayashi-Maskawa (CKM) matrix elements  and $\alpha$ is the electromagnetic coupling constant. The $\nu_i$ denotes the neutrinos with flavors $i=e,\mu,\tau$. The value of the Wilson coefficient $C_L$ within the SM was calculated at the next-to-leading order 
in QCD~\cite{Buchalla:1998ba, Misiak:1999yg}. It is given by $C_L=-X(x_t)/\sin^2\theta_W$, where $x_t=m_t^2/m_W^2$ and $X(x_i)$ is the 
corresponding loop function with $X(x_t) = 1.469\pm 0.017$~\cite{Girrbach-Noe:2014kea}. The  right-handed Wilson coefficient $C_R$ is negligible within the SM, but can
be induced in beyond the SM (BSM) scenarios. We therefore  keep the explicit dependence on this coefficient in the analytical expressions.

We denote the four-momenta of the $B$, $K^\ast$, $K$ and $\pi$ mesons  by $p_B, k, p_K$ and $p_\pi$, respectively, while the four-momenta of the neutrino and the antineutrino are denoted by $p_\nu$ and $p_{\bar{\nu}}$.  We use $q=p_\nu+p_{\bar{\nu}}$ and $p=p_K+p_\pi$.
$m_X$ denotes the mass of the meson $X=B,K^*,K,\pi$. The polarization vectors of the $K^\ast$-meson and the neutrino-pair in the rest frame of the $B$-meson are given in Appendix~\ref{App:Polarization vectors}.

\subsection{$B\to K^\ast \nu\bar{\nu}$}\label{Onshell}

We  recall the expressions for the $B\to K^\ast\nu\bar{\nu}$ decay amplitude and differential decay rate  for an asymptotic $K^*$-meson state.
The amplitude for $B\to K^\ast \nu_i\bar{\nu_i}$ decays, with  fixed $K^*$-polarization $n=\pm,0$, can be written as 
\begin{equation}
\mathcal{A}(n)=-\frac{4G_F}{\sqrt{2}}\lambda_t\frac{\alpha}{8\pi}h_\mu(n) \ell^\mu,
\end{equation}
where
\begin{equation}
h_\mu(n)=(C_L+C_R)\langle K^\ast(n)\vert \bar{s}\gamma_\mu b\vert B\rangle +(C_R-C_L)\langle K^\ast(n)\vert \bar{s}\gamma_\mu\gamma_5 b\vert B\rangle,
\end{equation}
and $\ell^\mu$ denotes the matrix element of the vector-minus-axial neutrino current between the vacuum and the neutrino pair. The matrix elements of the vector and axial-vector currents between the $B$ and $K^\ast$ mesons are parameterized in terms of the standard form factors, explicitly given in Appendix~\ref{App:B}. We use the $B\to K^\ast$ form factors given in Ref.~\cite{Straub:2015ica}, which were obtained from a combined fit \cite{Hambrock:2013zya} of  lattice QCD~\cite{Horgan:2015vla} and  light-cone sum rules (LCSR) results~\cite{Straub:2015ica}.

The hadronic amplitudes $h_\mu$ are written in terms of the hadronic helicity amplitudes $H$, which are defined as  projections of the hadronic matrix onto the polarization vectors of the neutrino pair for a given $K^\ast$-polarization $n$ as
\begin{equation}
H_{n}=\tilde{\epsilon}^{\mu\ast}_{n}\,h_\mu(n),\quad n=\pm,0.
\label{projection}
\end{equation}
For easier comparison with the literature, we switch to the transversity basis of perpendicular $(\perp)$ and parallel $(\parallel)$ polarizations via  $H_\perp=1/\sqrt{2}(H_+-H_-)$ and $H_\parallel=1/\sqrt{2}(H_++H_-)$, while the $H_0$ remains unchanged from  Eq.~\eqref{projection}.
The  hadronic transversity amplitudes  then read
\begin{equation}
\begin{split}
&H_\perp(q^2)=\frac{\sqrt{2}(C_L+C_R)\lambda^{1/2}(m_B^2,q^2,m_{K^\ast}^2)}{m_B+m_{K^\ast}}V(q^2),\\
&H_\parallel(q^2)=\sqrt{2}(C_L-C_R)(m_B+m_{K^\ast})A_1(q^2),\\
&H_0(q^2)=-\frac{1}{2m_{K^\ast}\sqrt{q^2}}(C_L-C_R)\bigg[(m_B+m_{K^\ast})(m_B^2-m_{K^\ast}^2-q^2)A_1(q^2)-\frac{\lambda(m_B^2,q^2,m_{K^\ast}^2)}{m_B+m_{K^\ast}}A_2(q^2)\bigg] \, ,
\end{split}
\end{equation}
where $\lambda(a, b, c) = a^2 + b^2 + c^2 -2 (ab+bc+ca)$.
The differential decay distribution in $q^2$, the square of the invariant mass of the $\nu\bar{\nu}$ pair, is then given as
\begin{eqnarray}\label{eq:dBrds:narrow}
\frac{d\Gamma}{dq^2}&=&3\frac{G_F^2\vert\lambda_t\vert^2\alpha^2\vert\vec{q}\vert q^2}{128\times 3 \pi^5m_B^2}\bigg[\vert H_\perp\vert^2+\vert H_\parallel\vert^2+\vert H_0\vert^2\bigg]\label{Br},\,
\end{eqnarray}
with  $\vert\vec{q}\vert=\lambda^{1/2}(m_B^2,m_{K^{\ast}}^2,q^2)/(2m_B)$. 
Here, the overall factor of three  comes from the summation over three flavors of the final state neutrinos.
Formula~\eqref{Br} agrees with the corresponding results in~\cite{Colangelo:1996ay, Altmannshofer:2009ma}. Integrating this distribution over the full kinematic region 
$0 \leq q^2 \leq (m_B-m_{K^*})^2$ we obtain for  the SM branching ratio  
\begin{align}
\mathcal{B}(B\to K^\ast\bar{\nu}\nu) = (9.49 \pm 1.01)\times 10^{-6} \, , 
\end{align} consistent with~\cite{Buchalla:2000sk,Buras:2014fpa, Girrbach-Noe:2014kea}. 
Partial branching ratios in low and high $q^2$-regions are given in Tab.~\ref{table:Br}; see Sec.~\ref{Numerical Analysis} for the definition of our current choice of binning in $q^2$. 
Separate $q^2$ regions are needed since form factors for the background modes $B\to K_0^* (\to K\pi)\nu \bar \nu$ and $B\to K\pi \nu \bar \nu$ are presently not available in the full $q^2$-region.

\section{Resonant contributions $B\to K_{res}(\to K\pi)\nu \bar \nu$}\label{scalars}

In this section we treat the $K^\ast$  at finite width and include  intermediate scalar states  decaying as well  to $K \pi$.
Such effects have been studied previously for $B\to K^\ast\ell^+\ell^-$ decays~\cite{Becirevic:2012dp,Matias:2012qz, Blake:2012mb}, and measured recently by the LHCb collaboration~\cite{Aaij:2016flj}.
In Sec.~\ref{sec:analytic} we obtain decay amplitudes and distributions for  $B \to (K^*,K_0^*,\kappa)( \to K \pi) \nu \bar \nu$ decays. In Sec.~\ref{Numerical Analysis}  we work out their phenomenology.

\subsection{Amplitudes and observables \label{sec:analytic}}

The total, resonant amplitude with fixed polarization $n$ of the final $K\pi$ pair can be written as
\begin{equation}
\small
\begin{split}
\mathcal{A}(B\to K_{res}&(n)(\to K\pi)\bar{\nu}_i\nu_i)=\\
=& -\frac{4G_F}{\sqrt{2}}\lambda_t\frac{\alpha}{8\pi} \sum_{res}\langle K\pi\vert K_{res}(n)\rangle\bigg[(C_L+C_R)\langle K_{res}(n)\vert \bar{s}\gamma_\mu b\vert B\rangle +(C_R-C_L)\langle K_{res}(n)\vert \bar{s}\gamma_\mu\gamma_5 b\vert B\rangle\bigg]\ell^\mu \widetilde{BW}_{res}(p^2),\label{Resonant-amplitude}
\end{split}
\end{equation}
where $p^2= (p_K+p_\pi)^2$ denotes the  square of the invariant mass of the $K\pi$-pair.
We parameterize the propagator of the intermediate vector $K^\ast$ resonance by a Breit-Wigner ansatz
\begin{align} 
\begin{split} \label{eq:BWKst}
&\widetilde{BW}_{K^\ast}(p^2)=\frac{1}{p^2-m_{K^\ast}^2+i m_{K^\ast} \Gamma_{K^\ast}} \, ,
\end{split}
\end{align}
where $\Gamma_{K^*}$ denotes the (constant) width of the $K^*$~\cite{PDG}.
In absence of finite width $B \to K^*$ form factors we employ the available narrow-width ones instead.

For  the broad scalar states we follow Ref.~\cite{Becirevic:2012dp} and include the contribution of the $K_0^\ast(800)\equiv \kappa$ that modifies the tail of the $K_0^\ast(1430)$ resonance in the $p^2$-region relevant to the $K^*$,
\begin{align} 
\begin{split}
&\widetilde{BW}_{\text{scalar}}(p^2)=-\frac{g_\kappa}{p^2-(m_\kappa-i\Gamma_\kappa/2)^2}+\frac{1}{p^2-(m_{K_0^\ast}-i\Gamma_{K_0^\ast}/2)^2}.
\label{BW}
\end{split}
\end{align}
We employ the mass and the width of the  scalar state $\kappa$  from Ref.~\cite{DescotesGenon:2006uk}, and the ranges  of the magnitude and argument of  $g_\kappa$
given in~\cite{Becirevic:2012dp}, which are  compatible with $D\to K^\ast\ell \nu$ spectra~\cite{Becirevic:2012dp, delAmoSanchez:2010fd};
see Tab.~\ref{tab:input} for a compilation of numerical input used in this work.
For alternative descriptions, see~\cite{Doring:2013wka}, and  \cite{Das:2014sra}. We checked explicitly that the model (\ref{BW}) is consistent
 with the measurements of the scalar fraction $F_S$ and the $\cos\theta_K$-distribution in $B\to K^\ast(\to K\pi) \mu^+ \mu^-$ decays~\cite{Aaij:2016flj} and  the $p^2$-distribution near the $K^\ast$~\cite{Aaij:2016kqt}. 
In the future experimental checks can be explicitly performed directly for the dineutrino mode by measuring the interference observable $b$, see~\eqref{thetaK-resonant}.
As for the $K^*$ we neglect a possible $p^2$-dependence in the decay widths.

The $K^*,K_0^* \to K \pi$  decay amplitudes  are expressed  in terms of the  couplings $g_{K^\ast K\pi}$ and $g_{K_0^\ast K\pi}$, defined as
\begin{equation}
\begin{split}
\langle K^i(p_K)\pi^j(p_\pi)\vert K^{\ast}(k,n)\rangle=c_{ij}(\epsilon_n\cdot p_K)g_{K^\ast K\pi},\quad
\langle K^i(p_K)\pi^j(p_\pi)\vert K_0^\ast(k)\rangle=c_{ij}g_{K_0^\ast K\pi} \, .\label{KpiAmplitudes}
\end{split}
\end{equation}
Here,  $c_{ij}$  denote isospin factors  that depend on the charges of the final state mesons, {\it  i.e.,} $\vert c_{+-}\vert=\sqrt{2}\vert c_{00}\vert =1$.
The magnitudes of the couplings can be obtained from  the corresponding decay rates using 
\begin{equation}
\begin{split}
\Gamma(K^\ast \to K^i\pi^j)=\frac{\vert c_{ij}\vert^2}{24\pi m_{K^\ast}^2}g^2_{K^\ast K\pi}\vert \vec{p}_K\vert^3,\,\quad \Gamma(K_0^\ast\to K^i\pi^j)=\frac{\vert c_{ij}\vert^2}{8 \pi m_{K_0^\ast}^2}g^2_{K_0^\ast K\pi}\vert \vec{p}_K\vert,\label{two-body-rates}
\end{split}
\end{equation} 
where $\vert \vec{p}_K\vert = \lambda^{1/2}(m_{K_{(0)}^\ast}^2,m_K^2,m_\pi^2)/2m_{K_{(0)}^\ast}$.
These couplings are important for the understanding of  nonperturbative  strong interactions; $g_{K^\ast K\pi}$ has been computed in  lattice QCD~\cite{Prelovsek:2013ela}, consistent with data~\cite{PDG}.

We write  the amplitude for the $K^\ast\to K\pi$ transition in the $K\pi$ rest frame using the components of the kaon's four-momentum, that is, for the coordinate system defined in Appendix~\ref{App:Polarization vectors},  given by $p_K^{\mu}=(E_K,0, \vert\vec{p}_K\vert\sin\theta_K, \vert\vec{p}_K\vert\cos\theta_K)$.  We defined $\theta_K$ as the angle between the kaon and the opposite direction of the $B$-meson in the $K\pi$ rest frame. The polarization vectors of the $K^\ast$ resonance in this frame are: $\epsilon_\pm^\mu=1/\sqrt{2}(0,\pm 1,i, 0)$ and $\epsilon_0^\mu=(0,0,0,1)$, resulting in $\epsilon_\pm\cdot p_K=-i\,\frac{1}{\sqrt{2}}\vert\vec{p}_K
\vert\sin\theta_K,\,\, \epsilon_0\cdot p_K=-\vert\vec{p}_K\vert\cos\theta_K.$

Using the projection~\eqref{projection} and the scalar form factor \eqref{scalar-form-factor}, we obtain the corresponding hadronic helicity amplitude
\begin{equation}
H_0'(q^2)=\,(C_R-C_L)\frac{\lambda^{1/2}(m_B^2,q^2,m_{K^\ast_0}^2)}{\sqrt{q^2}} f_+(q^2) \, .
\end{equation}
For the form factor $f_+(q^2)$ we use the results of the QCD-sum-rules computation from Ref.~\cite{Aliev:2007rq}, see Appendix~\ref{App:B}.

To combine vector and scalar resonance effects we write the differential decay rate for the four-body final state process by introducing the helicity amplitudes, distinguished by the \emph{tildae} labels, that incorporate, with the use of Eq.~\eqref{KpiAmplitudes}, the subsequent decay amplitude of the resonance into the final $K\pi$ pair, that is:
\begin{equation}
\begin{split}
&\widetilde{H}_{\parallel,\perp}(q^2,p^2,\cos\theta)=-i\frac{1}{\sqrt{2}} g_{K^\ast K\pi} \vert\vec{p'}_K\vert \sin\theta_K \,  \widetilde{BW}_{K^\ast}(p^2) \, H_{\parallel,\perp}(q^2),\\
&\widetilde{H}_0(q^2,p^2,\cos\theta)=-g_{K^\ast K\pi} \vert \vec{p'}_K\vert \cos\theta_K \, \widetilde{BW}_{K^\ast} (p^2) \, H_0(q^2),\\
&\widetilde{H}'_0(q^2,p^2)= g_{K_0^\ast K\pi} \widetilde{BW}_{\text{scalar}}(p^2) \,  H_0' (q^2),\label{Htilde}
\end{split}
\end{equation}
where $\vert \vec{p'}_K\vert$ is defined in Eq.~\eqref{factors} below.
Using the expression for the four-body phase space~{\it e.g.,} \cite{Das:2014sra,Becirevic:2016hea}, we obtain the three-fold decay distribution
\begin{eqnarray}
\begin{small}
\frac{d^3\Gamma}{dq^2 dp^2 d\cos\theta_K}=\frac{N(q^2)\vert\vec{q'}\vert \vert \vec{p'}_K\vert}{8 (2\pi)^5 m_B^2\sqrt{p^2}}\bigg[\sum_{i=\parallel,\perp,0}\vert \widetilde{H}_i\vert^2+2\Re(\widetilde{H}_0 \widetilde{H}'^\ast_0)+\vert \widetilde{H}'_0\vert^2\bigg]    \, , \label{resonant-formula}
\end{small}
\end{eqnarray}
with
\begin{align}
N(q^2)=G_F^2\lambda_t^2\alpha^2 q^2/(8\pi^2), \quad
\vert \vec{p'}_K\vert=\lambda^{1/2}(p^2,m_K^2,m_\pi^2)/(2\sqrt{p^2} )\, ,  \quad 
\vert\vec{q'}\vert=\lambda^{1/2}(m_B^2,p^2,q^2)/(2m_B) \, .\label{factors}
\end{align}
Eq.~\eqref{Br} is recovered in the NWA for the $K^\ast$ after setting the scalar contributions to zero and integrating the above distribution over $\cos\theta_K$ in the interval $(-1,1)$. 

The three-fold differential decay distribution~\eqref{resonant-formula} can be written as\begin{equation}
\frac{d^2\Gamma}{dq^2 dp^2 d\cos\theta_K}=a(q^2,p^2)+b(q^2,p^2)\cos\theta_K+c(q^2,p^2)\cos^2\theta_K.~\label{thetaK-resonant}
\end{equation}
where, schematically,
\begin{align} \nonumber
a(q^2,p^2)  \sim  \sum_{i=\parallel,\perp} \vert \vec{p}_K\vert^2 \frac{\vert {H}_i\vert^2}{2} + \vert {H}_0^\prime \vert^2   &\, ,  \quad  \quad 
 b(q^2,p^2) \cos \theta_K   \sim - 2 \vert \vec{p}_K\vert  \Re ( {H}_0 {H'}^\ast_0)    \, ,  \\
 c(q^2,p^2)  \cos^2 \theta_K  & \sim     \vert \vec{p}_K\vert^2   \left(    \vert H_0 \vert^2      -          \sum_{i=\parallel,\perp} \frac{ \vert {H}_i\vert^2 }{2}      \right)    \, . 
\end{align}
The parameterization~\eqref{thetaK-resonant} is general for contributions from spin 0 and spin 1 kaon resonances. For spin $\geq 2$ further powers of $\cos \theta_K$ arise.
The coefficient functions $a(q^2,p^2),b(q^2,p^2),c(q^2,p^2)$  represent  the three independent observables  that can be measured in angular analysis in $\theta_K$.
Instead of $a,b,c$ for phenomenology we consider 
the $q^2$-differential decay rate~\footnote{In what follows,  a single argument implies that the other variable has been integrated over, {\it e.g}., $b(q^2)=\int d p^2 b(q^2,p^2)$  and $b(p^2)=\int d q^2 b(q^2,p^2)$ etc.},
\begin{align}
\frac{d\Gamma}{dq^2}=2 \left( a(q^2)+ \frac{c(q^2) }{3}\right) \, , 
\end{align}
and the  longitudinal polarization fraction of the vector meson, $F_L$~\cite{Altmannshofer:2009ma},
\begin{equation}
F_L=\frac{d\Gamma_L/dq^2}{d\Gamma/dq^2},\quad     \frac{d\Gamma_L}{dq^2}=\frac{2}{3}  \left(a(q^2)+c(q^2) \right)  \, , \label{FL-resonant}
\end{equation}
both obtained after integration over $p^2$.
As usual, the $q^2$-averaged (binned) versions of  ratio-type observables are defined as
\begin{equation} \label{FLbinned}
\langle F_L\rangle =\frac{\Gamma_L}{\Gamma} \, , \quad \quad \Gamma_{(L)}=\int_{q^2_\text{min}}^{q^2_\text{max}} \frac{d\Gamma_{(L)}}{dq^2} \, . 
\end{equation}
Note that  $F_L$ does not depend on the Wilson coefficients if right-handed currents can be neglected. In this case, which includes the SM and which may be checked elsewhere, $F_L$ is probing form factors in the entire $q^2$-region. With a single observable it is not possible to extract  two form factor ratios without further input.
This is different in  $B \to K^* \ell^+ \ell^-$ decays at high $q^2$ which allows for a fit~\cite{Hambrock:2012dg}.
 Within the NWA for the $K^\ast$ we find after integration over the full $q^2$-region 
\begin{equation} 
\langle F_L\rangle_{\rm NWA} =0.49\pm 0.04 \,,
\end{equation}
consistent with Ref.~\cite{Altmannshofer:2009ma}.

In addition, we consider the  forward-backward asymmetry $A_{\rm FB}^K$, or alternatively, $A_{\rm FB \, L }^K$,
\begin{align} \label{eq:AFB}
A_{\rm FB \, (L)}^K \equiv  \frac{    \int_0^1 d \cos \theta_K   \frac{d^2\Gamma}{dq^2 d \cos \theta_K}  -   \int_{-1}^0 d \cos \theta_K   \frac{d^2\Gamma}{dq^2 d \cos \theta_K} }{\Gamma_{(L)}} =\frac{b(q^2,p^2)}{\Gamma_{(L)}} \,,
\end{align}
induced by  interference of the $K^\ast$ with intermediate scalar states. It  can be used to further check  the size of the scalar background, as  pointed out for $B \to K^* \ell^+ \ell^-$ decays in Ref.~\cite{Becirevic:2012dp}. By the same argument, $b=0$ in the presence of  vector $K^\ast$ only. Note that  contributions from $b(q^2,p^2)$ disappear 
from~\eqref{thetaK-resonant} after symmetric  $\cos \theta_K$-integration.
In the same way as in $F_L$ the dependence on Wilson coefficients drops out in $A_{\rm FB}^K$ if right-handed currents are negligible.
On the other hand, in $A_{\rm FB \, L }^K$ only amplitudes with $C_L-C_R$ enter, so the Wilson coefficients cancel in this ratio model-independently.

\subsection{Numerical Analysis}\label{Numerical Analysis}

We employ two different integration regions for $p^2$~\cite{Das:2014sra, Aaij:2016flj}
\begin{align} \nonumber
[(m_{K^\ast}-0.1\,\text{GeV})^2,  &(m_{K^\ast}+0.1\,\text{GeV})^2]  \, \quad \quad & \mbox{P-cut} \\
[(m_{K}+m_\pi)^2,  & 1.44\,\text{GeV}^2]  \, \quad \quad & \mbox{(S+P)-cut}     \label{eq:p2cuts}
\end{align}
where the first  one refers to the $K^*$ signal region and the second one  to a wider one, that allows to study backgrounds.

After fixing the integration limits for $p^2$, the endpoint in $q^2$ is a function of $p^2$, that is $q^2_{\text{max}}=(m_B-\sqrt{p^2})^2$. Note that some care is required with the comparison of the experimental results that follow from some choice of finite integration region in $p^2$ with the result of Eq.~\eqref{Br}. If one assumes the Breit-Wigner type parametrizations, as above, and applies the chosen $p^2$-cut, the differential decay rates over $q^2$ and the resulting total rates are smaller than those from  Eq.~\eqref{Br}. This can be explicitly seen from Tab.~\ref{table:Br} and has also been pointed in Ref.~\cite{Meissner:2013pba} for  $B_s\to K^\ast \ell\bar \nu$ decays. 

\begin{figure}[H]
\centering{
\includegraphics[height=0.3\textwidth]{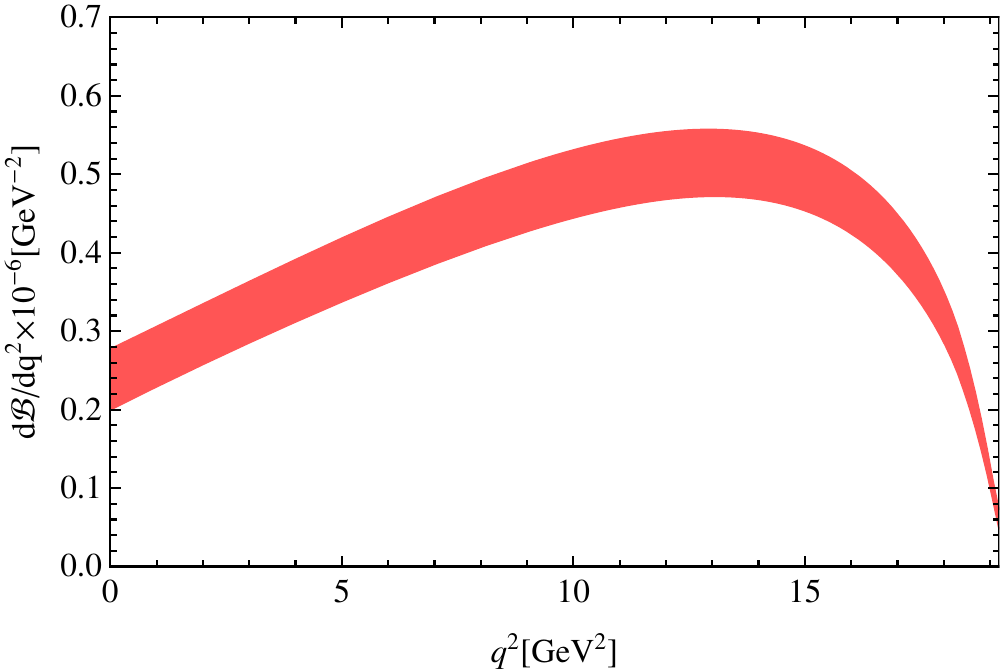}
\includegraphics[height=0.3\textwidth]{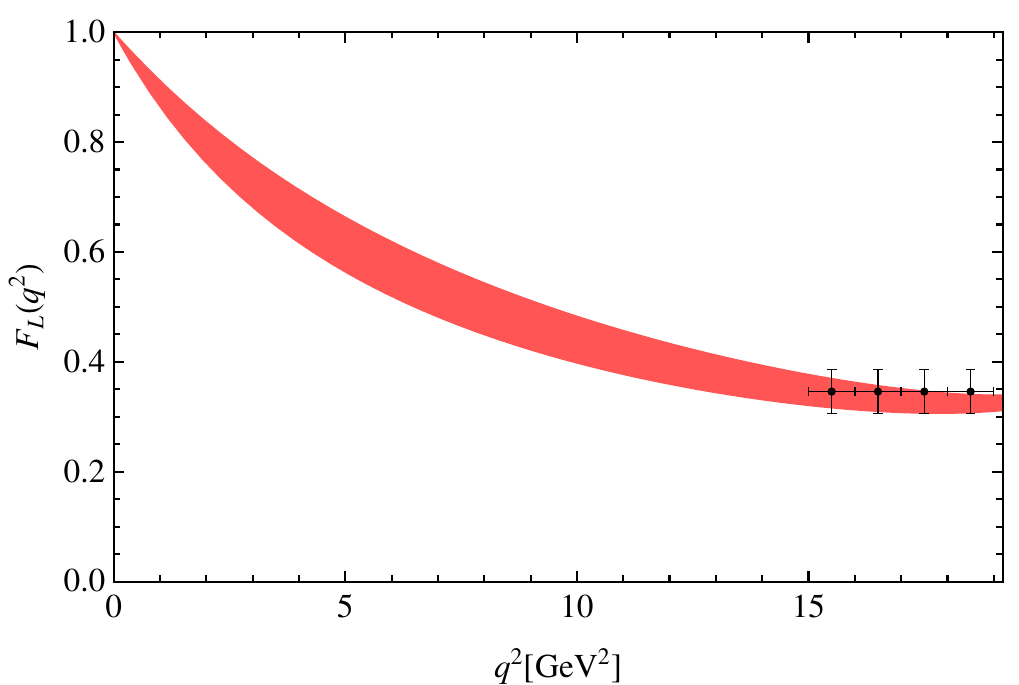}
}
\caption{\small Shown on the left is the differential SM branching fraction for the decay $B\to K^\ast(\to K\pi)\nu\bar{\nu}$ as the function of $q^2$, integrated over  $p^2$ within the P-cut, see~\eqref{eq:p2cuts}.
Shown on the right is the longitudinal polarization fraction $F_L(q^2)$.  Here, the (black) data points correspond to  form factor computations from
lattice QCD~\cite{Horgan:2015vla, Horgan:2013hoa}.
The error bands result from the uncertainties in the form factors taken from a combined fit \cite{Straub:2015ica}, and parametric inputs.} \label{fig:dBdq2892}
\end{figure}

We begin with the pure $B\to K^\ast(\to K\pi)\nu\bar{\nu}$ decays  at finite width. In Fig.~\ref{fig:dBdq2892} we show the  branching ratio and $F_L$ in the SM as functions of $q^2$ in the signal (P-cut) window. Note that $F_L$ goes to 1 and $1/3$ at maximal and zero recoil, respectively, as  dictated by helicity.  
On top of the uncertainty bands from $B \to K^*$ form factors
and parametric inputs we show for $F_L$ exemplarily predictions from lattice form factors~\cite{Horgan:2015vla, Horgan:2013hoa}. Recall that $F_L$ is unaffected by BSM physics if $C_R$ is negligible. As we will show, $F_L$ in addition receives only small uncertainties from scalar and non-resonant backgrounds. We therefore suggest it as a probe of form factor
calculations in the full $q^2$-region.

\begin{figure}[H]
\centering{
\includegraphics[height=0.30\textwidth]{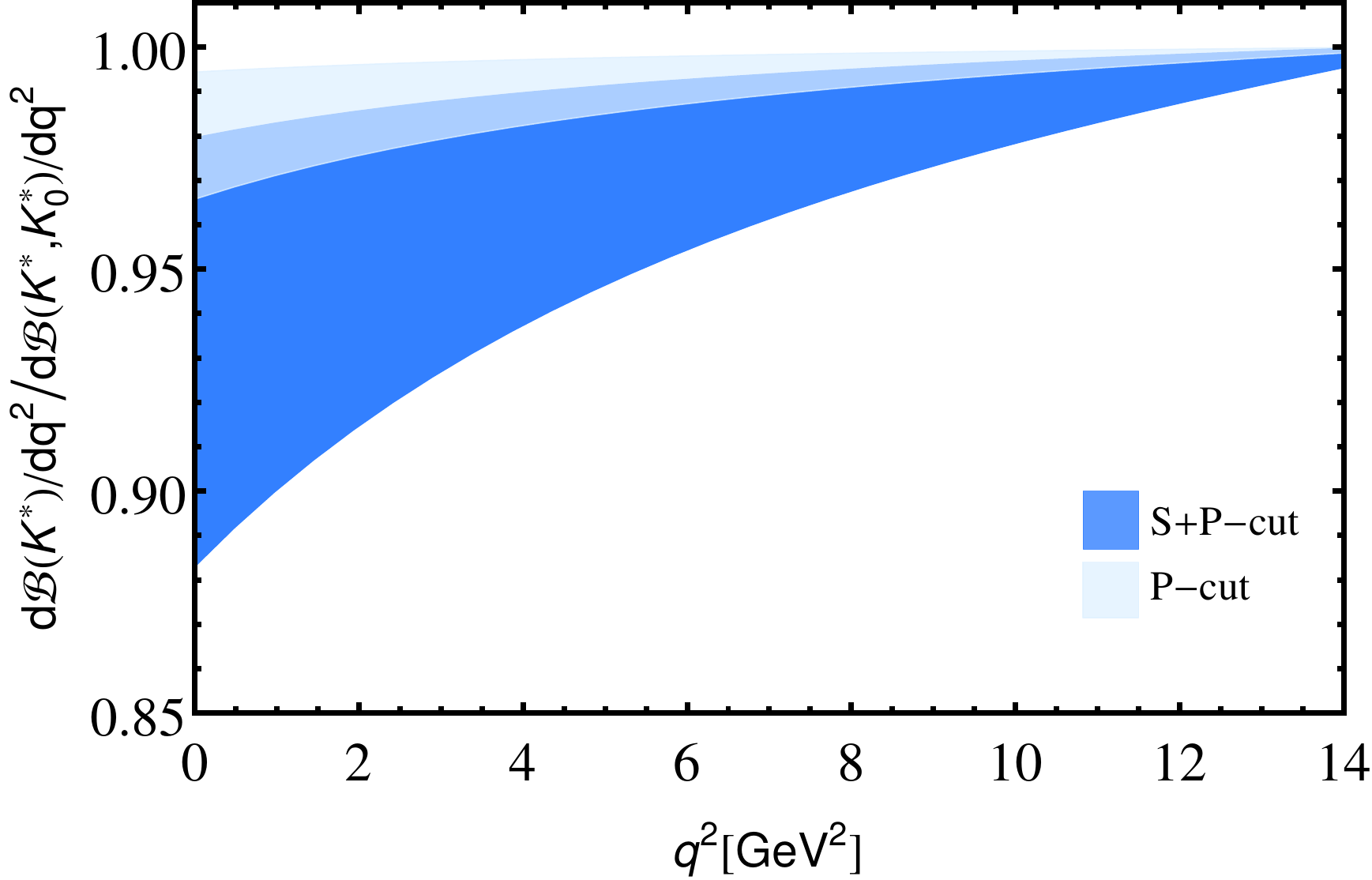}
}
\caption{\small The ratio  of  $B\to K^\ast(\to K\pi)\nu\bar{\nu}$  to  $B \to (K^*,K_0^*,\kappa)( \to K \pi) \nu \bar \nu$  $q^2$-differential branching ratios for different $p^2$-cuts.
We set the $B\to K^\ast$ form factors  and other inputs to their central values, except for the parameters which enter the scalar meson contributions ($B\to K_0^\ast$ form factors, $g_\kappa$ and other parameters in~\eqref{BW}). We leave these parameters free within the corresponding errors, such that the boundaries of the above bands correspond to the resulting minimal and maximal values.  
\label{fig:VoverVplusS}}
\end{figure}

In the following  it becomes  necessary to separate the analysis into two $q^2$-regions,
"low $q^2$" within the range $ [0 -14]\,\text{GeV}^2$ and "high $q^2$"
within $q^2  \in [14 -19]\,\text{GeV}^2$.
We refrain from presenting numerical predictions for 
intermediate scalar resonances  at high $q^2$, where the
extrapolation of the scalar form factors Eq.~\eqref{scalar-ff} into
the highly off-shell (for $K_0^\ast$) region is required.
Instead, we find these effects  to be highly sub-dominant in this region,
their kinematic  suppression towards the high $q^2$-region is evident from  Fig.~\ref{fig:VoverVplusS}.
The other  reason for a separation in $q^2$ are the non-resonant effects, whose  description using chiral methods is expected to hold only at  high $q^2$, see Sec.~\ref{Non-resonant contributions}. The estimate of non-resonant effects in the low $q^2$-region is beyond the scope of this work.

\begin{figure}[htb]
\centering{
\includegraphics[height=0.33\textwidth]{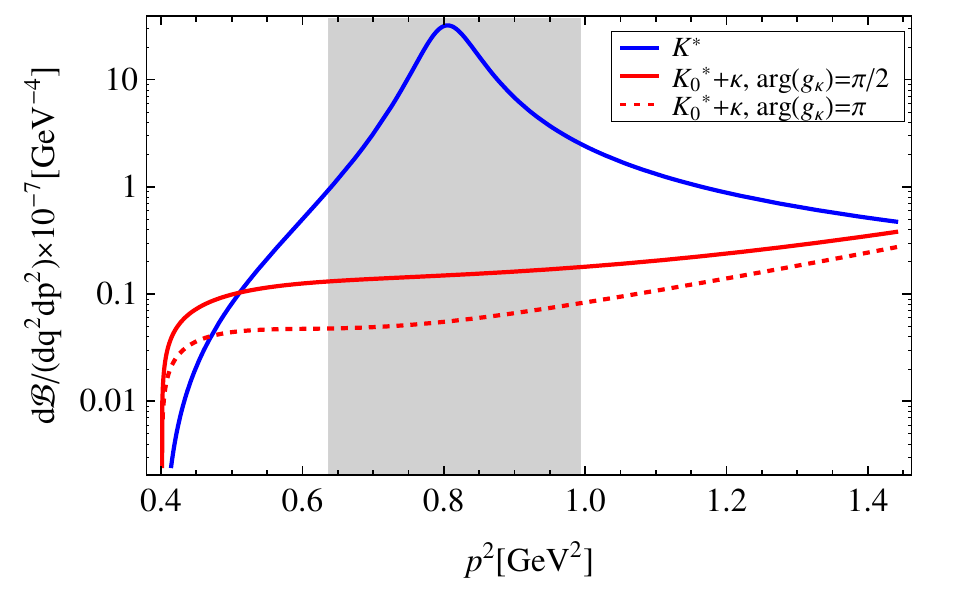}
}
\caption{\small 
Lineshapes of the resonant $K^\ast$ (\ref{eq:BWKst}) (solid blue line) and scalar mesons ($K_0^\ast(1430), \kappa(800)$) (red line) at $q^2=4\,\text{GeV}^2$.  All input parameters are set to their central values, and $g_\kappa=0.2,  \arg(g_\kappa)=\pi/2$ (red solid) and  $\arg(g_\kappa)=\pi$ (red dashed), see Eq.~\eqref{BW}. 
The (gray) shaded region corresponds to the P-cut in $p^2$, cf. \eqref{eq:p2cuts}.  } \label{Lineshapes}
\end{figure}

The  lineshapes of the $K^*$ and the scalars are shown in Fig.~\ref{Lineshapes}. They do not interfere in the $B \to (K^*,K_0^*,\kappa)( \to K \pi) \nu \bar \nu$ differential branching ratio. Contributions from scalars underneath the $K^*$ peak can be probed with side-band measurements or $A_{\rm FB \, (L)}^K$ (\ref{eq:AFB}).

The impact  of the intermediate scalar meson states on $B\to K^\ast(\to K\pi)\nu\bar{\nu}$ is illustrated in Fig.~\ref{fig:VoverVplusS}.
Shown  is the ratio of  $B\to K^\ast(\to K\pi)\nu\bar{\nu}$  to  $B \to (K^*,K_0^*,\kappa)( \to K \pi) \nu \bar \nu$  differential branching ratios~\footnote{This ratio corresponds to $1-F_S$, where $F_S$ denotes the fraction of scalar contributions \cite{Blake:2012mb}.} for different $p^2$-cuts.
 After integrating over  $\cos\theta_K$, the information on the interference between the scalar and vector amplitudes is lost and the corresponding branching fractions can simply be added.  We find that the impact of  the scalars drops with increasing $q^2$. It is at most $\sim 10 \%$ for the S+P-cut and few percent in the P-cut.

After integrating over the  low $q^2$-region the ratio between the corresponding integrated rates deviates from  unity at the level of at most $\sim 1\%$ in the P-cut and $\sim4\%$ for the (S+P)-cut. Branching ratios involving  only the $K^*$, and only the scalars integrated over the low $q^2$-region are  given in Tab.~\ref{table:Br}.
The ranges given for the scalars correspond to the minimal and maximal values obtained from the ranges of the scalar form factors (see Eq.~\eqref{scalar-ff} and the text below), and parameters in~\eqref{BW}.

 \begin{table}[H]
\centering
\renewcommand{\arraystretch}{1.6}
 \begin{tabular}{l  c c c c c}\hline \hline
  &   low $ q^2 \in  [0 -14]\,\text{GeV}^2$~ &  high $q^2  \in [14 -19]\,\text{GeV}^2$ \\\hline
 $\mathcal{B}(B\to K^\ast \nu\bar{\nu})|_{\text{NWA}} $   & ~$6.96\pm 0.76$ & ~$2.50\pm 0.22$ \\                                
 \hline
  $\mathcal{B}(B\to K^\ast(\to K\pi)\nu\bar{\nu})|_{\text{P-cut}}$  & ~$6.01\pm 0.65$ & ~$2.09\pm 0.22$  \\
   $\mathcal{B}(B\to K^\ast(\to K\pi)\nu\bar{\nu})|_{\text{S+P-cut}}$  & ~$6.80\pm 0.73$ & ~$2.29\pm 0.23$ \\ 
  \hline
  $\mathcal{B}(B \to (\kappa, K_0^\ast)(\to K\pi)\nu\bar{\nu})\big\vert_{\text{P-cut}}$ & 
  $[0.01 \ldots 0.07]$ &   ~$-$  \\
  $\mathcal{B}(B\to (\kappa, K_0^\ast)(\to K\pi)\nu\bar{\nu})\big\vert_{\text{S+P-cut}}$ & 
  $[ 0.04 \ldots 0.30]$ &    ~$-$ \\
    \hline
 $\mathcal{B}(B\to (K^\ast+\text{nonres})(\to K\pi)\nu\bar{\nu})|_{\text{P-cut}}$  & ~$-$ & ~$2.09\pm 0.22^{+0.42}_{-0.29}$ \\ 
 $\mathcal{B}(B\to (K^\ast+\text{nonres})(\to K\pi)\nu\bar{\nu})|_{\text{S+P-cut}}$  & ~$-$ & ~$2.29\pm 0.23^{+0.62}_{-0.27}$ \\    
\hline\hline
 \end{tabular}\caption{SM branching fractions in units of $10^{-6}$ for different cuts in $p^2$, see \eqref{eq:p2cuts}, and $q^2$ as indicated.
The (first) row with $B\to K^\ast\bar{\nu}\nu$ corresponds to the NWA, while in the second and third row  finite width effects (\ref{eq:BWKst}) of the $K^*$ have been included.
 The ranges given for the scalar resonance contributions correspond to the ranges of the scalar form factors (\ref{scalar-ff}) and the parameters in~\eqref{BW}. Interference between the 
 scalar- and vector-meson induced amplitudes is lost upon $\cos \theta_K$-integration such that the corresponding branching fractions can simply be added. 
 The second uncertainty in the last two rows stems mostly from the unknown strong phase $\delta$.
 The symbol  $-$ indicates that theoretical predictions are not available, see text for details. \label{table:Br}}
\end{table} 
For $F_L$ we find that its value for pure $B\to K^\ast(\to K\pi)\nu\bar{\nu}$ decays does not differ between NWA, the P- and (S+P)-cut predictions at finite width.
This can be expected since the $p^2$-dependence is universal for all $\widetilde{H}_i$, see  Eq.~\eqref{Htilde}. 
We also find that the effect of the scalar states on  $F_L$ is negligible compared  to other sources of uncertainties. SM values for $F_L$ are given in
Tab.~\ref{table:FLs}.
 \begin{table}[H]
\centering
\renewcommand{\arraystretch}{1.6}
 \begin{tabular}{l  c c c c c}\hline \hline
  &   low  $ q^2 \in  [0 -14]\,\text{GeV}^2$~ &  high $q^2  \in [14 -19]\,\text{GeV}^2$ \\\hline
 $\langle F_L\rangle|_{\text{NWA,P-, (S+P)-cut}}$   & ~$0.54\pm 0.04$ & ~$0.34\pm 0.02$ \\                                
    \hline
     $\langle F_L\rangle (B\to (K^\ast+\text{nonres})(\to K\pi)\nu\bar{\nu})|_{\text{P-, (S+P)-cut}}$  & ~$-$ & ~$0.34\pm 0.02 \pm 0.01$ \\ 
\hline\hline
 \end{tabular}\caption{$\langle F_L\rangle$  in the SM for different cuts in $p^2$, see \eqref{eq:p2cuts}, and $q^2$-binning as indicated, see Tab.~\ref{table:Br}. The entries in the first row are indistinguishable between the NWA  and the  finite width treatment  (\ref{eq:BWKst})  with P- and (S+P)-cuts. The impact of  scalar mesons is negligible. The last  row
 gives  $\langle F_L\rangle$  including non-resonant contributions.  
  \label{table:FLs}}
 \end{table}

\begin{figure}[H]
\centering{
\includegraphics[height=0.30\textwidth]{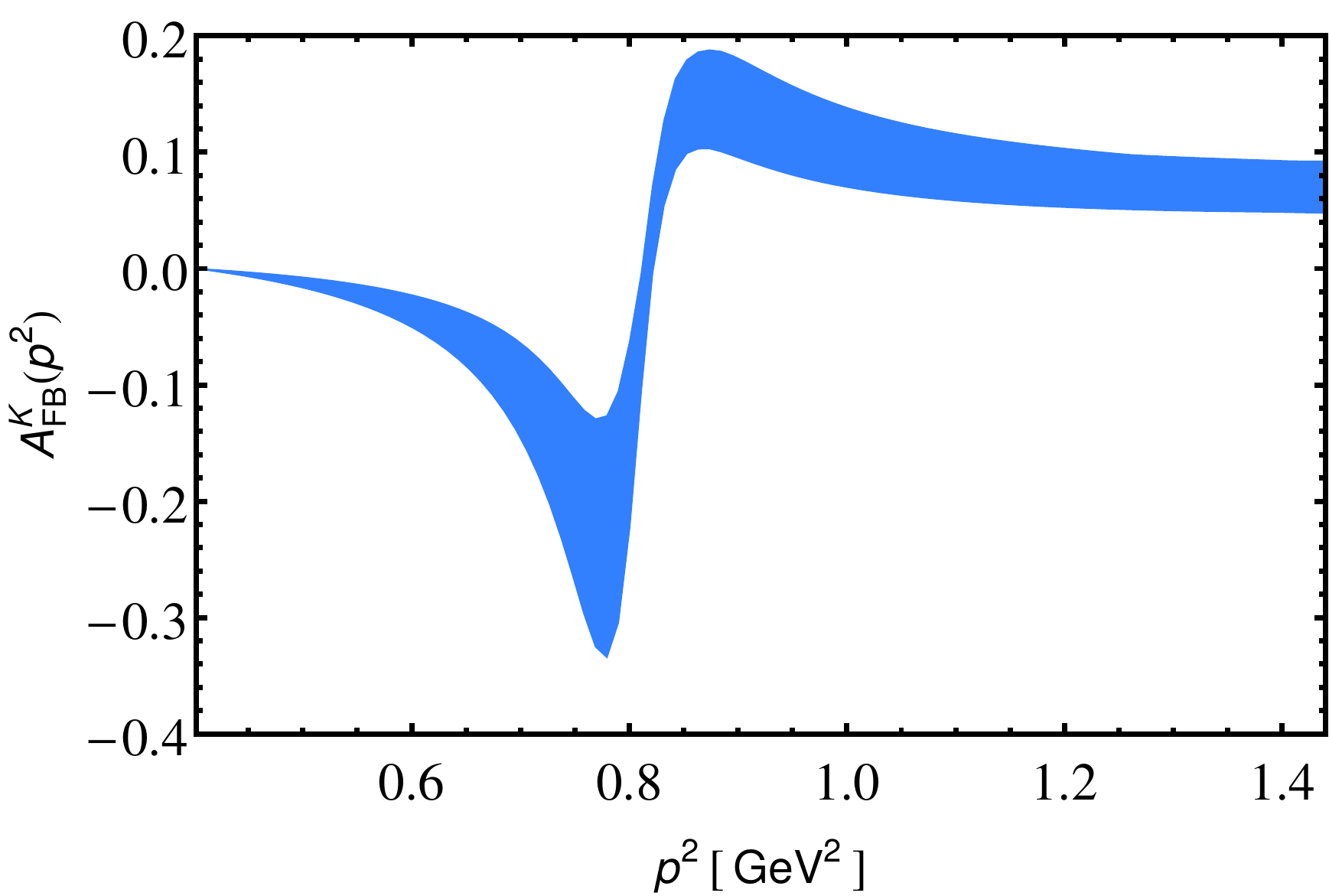}
\includegraphics[height=0.30\textwidth]{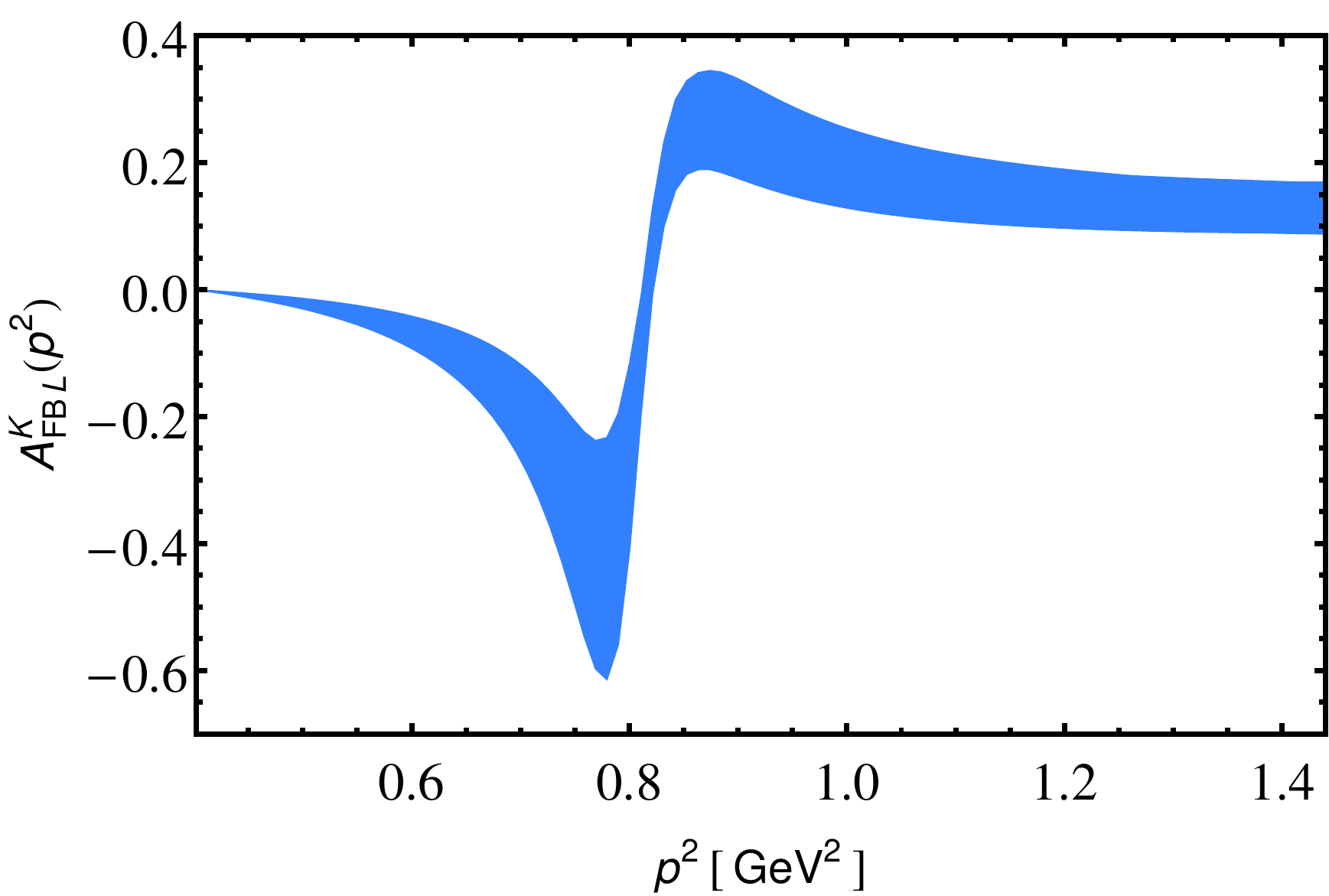}
}
\caption{The forward-backward asymmetry $A^K_{\rm FB}(p^2)$ (plot to the left)  and $A^K_{\rm FB \, L}(p^2)$ (plot to the right) defined in (\ref{eq:AFB}) 
integrated over the low $q^2$-region. The  corresponding decay rates have been  integrated over the low $q^2$-region and the (S+P)-cut in $p^2$.}\label{pb2}
\end{figure}
In Fig.~\ref{pb2} we show the forward-backward asymmetry  integrated over the low-$q^2$ region and normalized to the total rate in this region (left plot) and the total longitudinal rate (right plot), both integrated over $p^2$ in the (S+P)-cut. These observables can be used to test the model of the scalar contributions. The normalization to the longitudinal rate is particularly useful, since the Wilson coefficients $C_{L,R}$ drop out.  $A^K_{\rm FB \, (L)}(p^2)$ change sign across $p^2$. Therefore,
cancellations arise from integration over $p^2$ resulting in small values of  $A^K_{\rm FB \, (L)}(q^2)$, of the order of a percent.

\section{Non-resonant contributions}\label{Non-resonant contributions}

We consider non-resonant contributions to the four-body final state decay process.
These contributions were  studied in~\cite{Lee:1992ih} and were recently taken into account in  $B\to K^\ast\ell\ell$ decays in~\cite{Das:2014sra,Das:2015pna}, and some time ago in $D\to K\pi \ell\nu$ processes in~\cite{Bajc:1997nx}. 
The non-resonant matrix elements of the (axial-) vector currents between the $B$ and the  $K\pi$  can be parameterized 
as follows~\cite{Lee:1992ih} 
\begin{equation}
\begin{split}
&\langle K^i\pi^j\vert \bar{s}\gamma_\mu b\vert B\rangle=c_{ij} h\,\epsilon_{\mu\nu\alpha\beta}p_B^\nu(p_K^\alpha+p_\pi^\alpha)(p_K^\beta-p_\pi^\beta),\\
& \langle K\pi\vert \bar{s}\gamma_\mu\gamma_5 b\vert B\rangle=c_{ij}\big[-i\,w_+(p_{K\mu}+p_{\pi\mu})-i w_-(p_{K\mu}-p_{\pi\mu})-i r q_\mu\big] \, , \label{nr-matrix-elements}
\end{split}
\end{equation}
where the form factors $w_\pm, h$ and $r$ are functions of $q^2, p^2$ and $\theta_K$.
They are presently not known from  first-principles of QCD. We use the leading 
order Heavy-Hadron-Chiral-Perturbation-Theory (HH$\chi$PT) results~\cite{Lee:1992ih, Buchalla:1998mt} 
\begin{equation}
\begin{split}
&w_\pm(q^2,p^2,\theta_K)=\pm\frac{g f_{B_d}}{2f^2}\frac{m_B}{v\cdot p_\pi+\Delta},\\
&h(q^2,p^2,\theta_K)=\frac{g^2f_{B_d}}{2f^2}\frac{1}{(v\cdot p_\pi+\Delta)(v\cdot p_{K\pi}+\Delta+\mu_s)},\label{nr-ffs}
\end{split}
\end{equation}
where $v$ denotes the four-velocity of the $B$-meson, $f^2=f_\pi f_K$, $\Delta=m_{B^\ast}-m_B$ and $\mu_s=m_{B_s}-m_B$. We collect the corresponding numerical values of the inputs in Tab.~\ref{tab:input}. The results in~\eqref{nr-ffs} are expected to be valid only in the kinematic range in which chiral perturbation theory applies. This corresponds to  $p_B\cdot p_{\pi,K}/m_B\lesssim 1\,\text{GeV}$,  which is roughly satisfied in the high $q^2$-region.

The non-resonant $B \to K \pi \nu_i \bar \nu_i$ decay amplitude can be written as
\begin{equation}
\mathcal{A}(B\to K\pi\nu_i\bar{\nu}_i)=-\frac{4G_F}{\sqrt{2}}\lambda_t\frac{\alpha}{8\pi}\bigg[ (C_L+C_R)\langle K\pi\vert \bar{s}\gamma_\mu b\vert B\rangle +(C_R-C_L)\langle K\pi\vert \bar{s}\gamma_\mu\gamma_5 b\vert B\rangle\bigg]\ell^\mu.
\end{equation}

The non-resonant hadronic transversity amplitudes are obtained by projecting the matrix elements~\eqref{nr-matrix-elements} onto the polarization vectors of the neutrino pair, see Eq.~\eqref{projection}
\begin{align}
 H_\perp^{\text{nr}} &= (C_L+C_R)\sin\theta_K \frac{\lambda^{1/2}(m_{K\pi}^2,m_K^2,m_\pi^2)\lambda^{1/2}(m_B^2, p^2, q^2)}{2\sqrt{p^2}} h,\nonumber\\ 
H_\parallel^{\text{nr}}&=-(C_L-C_R)\sin\theta_K \frac{\lambda^{1/2}(p^2,m_K^2,m_\pi^2)}{\sqrt{p^2}}w_-,\\ 
H_0^{\text{nr}}&=\frac{i(C_L-C_R)}{2\sqrt{q^2}}\bigg[w_-\frac{1}{p^2}\bigg((m_K^2-m_\pi^2)\lambda^{1/2}(m_B^2,q^2,p^2)-(m_B^2-p^2-q^2)\lambda^{1/2}(m_K^2,m_\pi^2,p^2)\cos\theta_K\Big)\nonumber\\ 
&+w_+\lambda^{1/2}(m_B^2,q^2,p^2)\bigg].\nonumber
\end{align}
We model the three-fold differential decay distribution including resonance and non-resonance contributions as follows
\begin{eqnarray}
\small
\frac{d^3\Gamma}{dq^2 dp^2 d\cos\theta_K}=\frac{N(q^2)\vert\vec{q'}\vert \vert \vec{p'}_K\vert}{8 (2\pi)^5 m_B^2\sqrt{p^2}}\bigg[\vert e^{-i\,\delta}\widetilde{H}_\perp+H^{\text{nr}}_\perp\vert ^2+\vert e^{-i\,\delta}\widetilde{H}_\parallel+H^{\text{nr}}_\parallel\vert ^2+\vert e^{-i\,\delta}\widetilde{H}_0+H^{\text{nr}}_0+e^{-i\,\delta}\widetilde{H}'_0\vert^2\bigg] .\label{full-formula}
\end{eqnarray}
Here, we included $\delta$, a relative strong phase. There is just a single phase for all transversity amplitudes because all individual form factors can be chosen real-valued and 
by  approximate universality of the low recoil region.
In view of other uncertainties we do not consider $\delta$ depending on $q^2$, because one expects the phase to only slowly vary with $q^2$.
However, $\delta$ should vary with $p^2$. In neglecting this effect, which, in principle, could be taken care of, the strong phase becomes an effective  $p^2$-bin averaged phase.

\begin{figure}[htb]
\centering{
\includegraphics[height=0.33\textwidth]{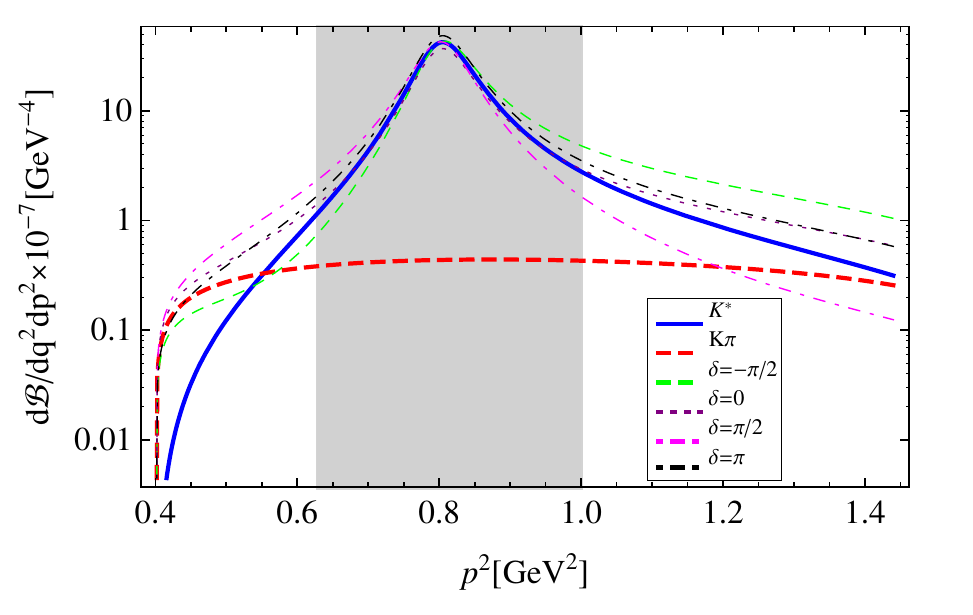}
}
\caption{\small 
Lineshapes of the resonant $K^\ast$-contribution (solid blue line), purely non-resonant contribution (dashed red line) and the  lineshapes which also include  interference effects for different values of the strong phase $\delta=0, \pm \pi/2,\pi$ at $q^2=16\,\text{GeV}^2$. The (gray) shaded region corresponds to the P-cut in $p^2$, cf. \eqref{eq:p2cuts}.} \label{Lineshapesnr}
\end{figure}

We stress that \eqref{full-formula} is a model, with model parameter $\delta$. Alternative descriptions would include modified Breit-Wigner propagators for the $K^*$, and
$B \to K^*$ form factors that take into account finite width effects.
The model \eqref{full-formula} can be  improved by data, for instance, by  measurements of  the lineshape  outside the $K^*$ signal region, see Fig.~\ref{Lineshapesnr}.

Furthermore, $\delta$ can be constrained from  measurements of the ratios of
angular coefficients in the process $B\to K^\ast\ell^+ \ell^-$, {\it e.g.,} $I_7/I_5$, $I_7/I_6$, see Eq.(14) in~\cite{Das:2015pna} for the complete list and more details. These angular coefficients are not observable in the dineutrino mode.
The benefit of using these ratios, as opposed to the total rate, lies in their independence on short-distance physics in the limit in which the right-handed operator can be neglected. 
Note that the angular coefficients become sensitive to the relative strong phase only in the $p^2$-region below and above the signal region (P-cut), see~\cite{Das:2015pna}.
Data on these coefficients exist  at present  only for  the signal region~\cite{Aaij:2015oid}.

The non-resonant amplitudes can be expanded in terms of orthonormal functions of the angle $\theta_K$, resulting in a distribution that is more complicated than~\eqref{thetaK-resonant}, which  arises solely from  vector and scalar meson states. Once higher waves $\ell \geq2$ are present the definition of the angular observables $F_L$  and $A_{\rm FB \, (L)}^K$  becomes more subtle. Here we use the projections via associated Legendre polynomials $P^0_0=1, P^0_1=\cos\theta_K, P^0_2=1/2(3\cos^2\theta_K-1)$ as
\begin{align}
\frac{d{\Gamma}_L}{dq^2}=& \int_{-1}^1\frac{d^2\Gamma}{dq^2d\cos\theta_K}\bigg(\frac{1}{3} P^0_0+\frac{5}{3}P^0_2\bigg)d\cos\theta_K \, ,
\label{Gamma_L-tilde} \\
{b}(q^2,p^2)= &\int_{-1}^1\frac{d^2\Gamma}{dq^2dp^2d\cos\theta_K}\frac{3}{2}P^0_1\,d\cos\theta_K \, , \label{btilde}
\end{align}
from which $\langle {F}_L\rangle$ and $A_{\rm FB \, (L)}^K$ follow as in Eqs.~\eqref{FLbinned} and (\ref{eq:AFB}), respectively.
In the limit in which only $\ell=0$ and $\ell=1$ effects are accounted for, one can insert the distribution~\eqref{thetaK-resonant} into the above formula to recover~\eqref{FL-resonant}.
Consequently, $b$ probes predominantly S,P-wave interference.

To illustrate the effect of the non-resonant amplitudes we present in Fig.~\ref{fig:dBrPhase} the  contributions to   the branching fraction and $\langle{F}_L\rangle$, integrated over  the high $q^2$-region, as  functions of the strong phase and normalized to the  pure $K^*$-case. We find that the resulting uncertainty in  the branching fraction is significant and can reach up to $20\%$  in the P-cut, while in $\langle {F}_L\rangle$ it is smaller, at  most at the level of $2.5\%$. 
\begin{figure}[htb]
\centering{
\includegraphics[height=0.3\textwidth]{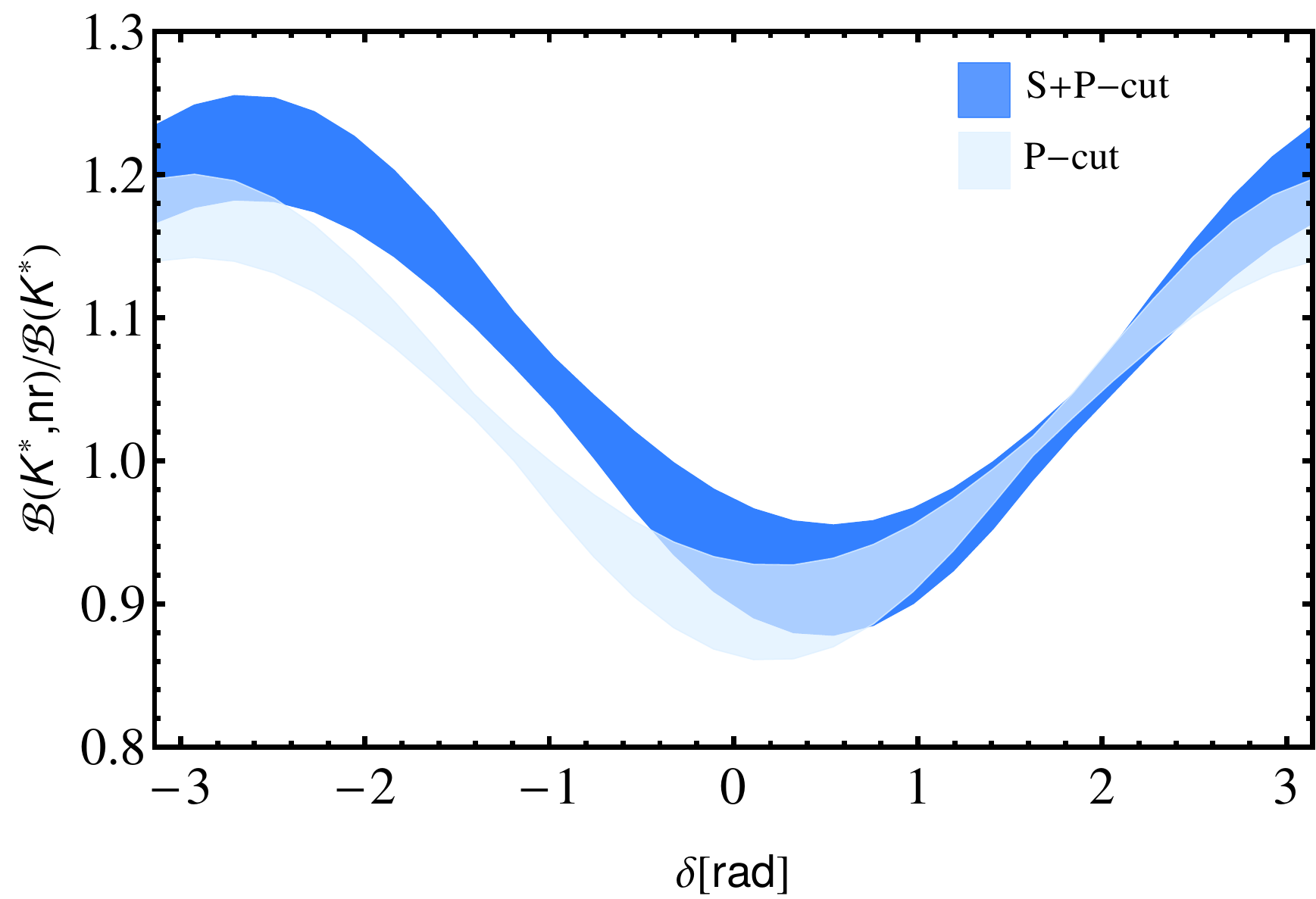}
\includegraphics[height=0.3\textwidth]{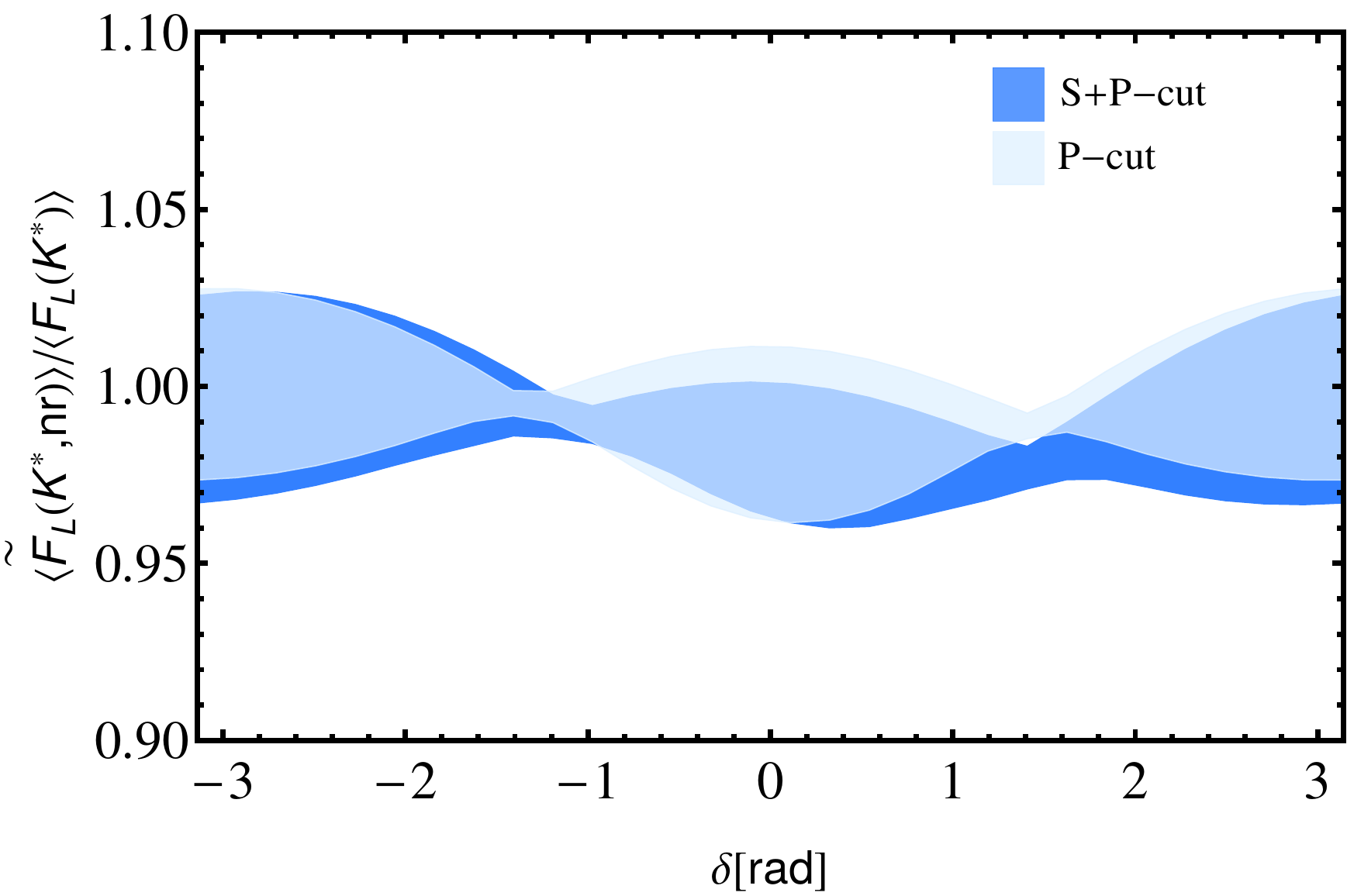}
}
\caption{\small Plot on the left: the ratio of $B\to (K^\ast + {\rm nr}) (\to K\pi)\nu\bar{\nu}$  to
$B\to K^\ast (\to K\pi)\nu\bar{\nu}$ in high- $q^2$-integrated  branching fractions
as a function of the relative strong phase $\delta$. Plot on the right: the same 
for  ${F}_L$.
We fixed the $B\to K^\ast$   form factors  and other parametric inputs to their
central values and varied the parameters in the non-resonant form factors via
uniform distributions. The boundaries of the bands correspond to minimal and maximal
values obtained in this way. The darker and lighter blue bands correspond to P- and 
(S+P)-cuts~\eqref{eq:p2cuts}, respectively.\label{fig:dBrPhase}
\label{fig:dBrPhase}}
\end{figure}

We quantify the effect of the non-resonant contributions at high $q^2$ on the branching ratio  in the last two rows of Tab.~\ref{table:Br}. The corresponding central value and the first errors are the same as the corresponding entries for the resonant contributions (two rows above). The second errors are the result of the variation of the parameters of the non-resonant form factors (including the strong phase $\delta$) via  uniform distributions, while keeping all other inputs fixed  to their central values. The upper and lower error represent the maximal and minimal distance from the central value. The corresponding predictions  for  $F_L$ are given in last row of Tab.~\ref{table:FLs}.

\begin{figure}[H]
\centering{
\includegraphics[height=0.32\textwidth]{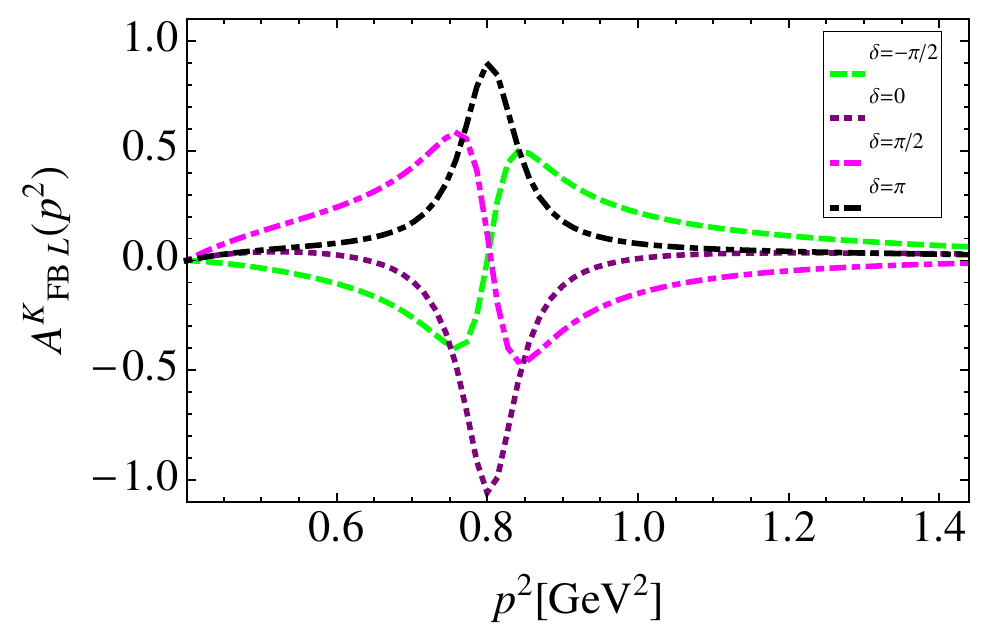}
\includegraphics[height=0.32\textwidth]{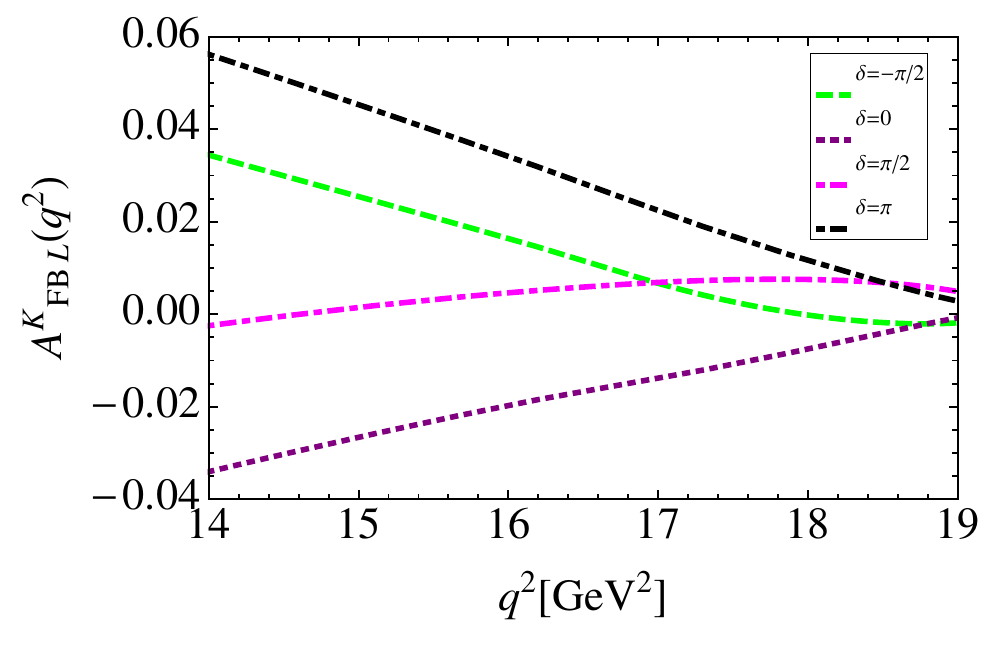}
}
\caption{The forward-backward asymmetry  $A_{\rm FB \, L}^K$ obtained using \eqref{btilde} as a function of $p^2$ after high $q^2$-integration (plot to the left) and as a function of $q^2$ after S+P-cut integration (plot to the right)
for different phases $\delta=0, \pm \pi/2, \pi$ and other input parameters fixed to central values. In each plot the
 longitudinal rate  \eqref{Gamma_L-tilde} is integrated over the  high $q^2$-region and over $p^2$ within the  (S+P)-cut  \eqref{eq:p2cuts}. }
\label{fig:btilde}
\end{figure}

In Fig.~\ref{fig:btilde} we show  $A_{\rm FB \, L}^K$ for different values of the strong phase $\delta$. Other numerical input is
fixed  to central values. 
While in $A_{\rm FB \, L}^K$ the Wilson coefficients $C_{L,R}$ drop out in the  $\ell=0,1$ limit, there is a residual dependence 
from the interference of the $K^\ast$ resonance with the D-wave components of the non-resonant amplitude. We checked that this effect is much smaller than the dependence on the strong phase $\delta$,
 which allows to experimentally constrain $\delta$.  Note that the corresponding uncertainties from the $B\to K^\ast$ form factors are between $5-8\%$, while the non-resonant form factors introduce additionally $\sim 15\%$. Already a rough determination of $\delta$   would significantly reduce the uncertainties in the $B \to K^* (\to K \pi)  \nu \bar \nu$  branching fraction, see Fig.~\ref{fig:dBrPhase}. 
 
Although presently there is no theory prediction available for non-resonant decays at low $q^2$,  $A_{\rm FB \, (L)}^K$ can be studied here experimentally as well  to constrain  the (non-resonant) backgrounds also in this region.
 
 \section{Conclusions   \label{sec:con} }
 
We revisited the decay $B \to K^* \nu \bar \nu$, its virtues and uncertainties. 
$B \to K^* \nu \bar \nu$ data on $F_L$ can be used, unlike   $B \to K^* \ell^+ \ell^-$ decays,  in the entire $q^2$-region to test $B \to K^*$ form factors from  lattice QCD  or other non-perturbative means in a model-independent way if right-handed currents can be neglected. 
The latter can be tested, for instance, using null-tests of the $B \to K^* \ell^+ \ell^-$ angular distribution.  The plot to the right in Fig.~\ref{fig:dBdq2892} 
illustrates the current level of form factor uncertainties. Probing form factors is limited by the resonant and non-resonant contributions considered in this  work. They need to be taken into account once the $K^*$ is treated beyond the narrow-width approximation. 

We analyzed {\it i}) finite width effects of the $K^*$, {\it ii)} the effects induced by non-resonant contributions in the region of low hadronic recoil (high $q^2$)
and {\it iii)} contributions from scalar resonances decaying to $K \pi$ at low $q^2$. The  restrictions to the kinematic regions in  {\it ii)}  and {\it iii)} originate from
the current availability of  $B \to K \pi$ and $B \to (K_0^*,\kappa)$ form factors, respectively. 

Our findings are summarized in Tabs.~\ref{table:Br}-\ref{table:FLs}. 
At high $q^2$  the non-resonant contributions in the model \eqref{full-formula} introduce an uncertainty of   $\mathcal{O}(0.1)$ in the branching ratio, and at around few $\%$ in the longitudinal polarization fraction $F_L$.  It is  desirable to check these  predominantly interference-induced effects  with either a global analysis of rare decay data
or further theoretical study.
On the other hand, the contributions of the scalar resonances  to the branching ratio in the low $q^2$-region are small with respect to other sources of uncertainty, and are at most of the order of $1\%$ in the signal region (P-cut) and $\lesssim 4\%$  for  the wider $S+P$-cut. Their effect on 
$F_L$ is negligible compared  to the uncertainties from the $B\to K^\ast$ form factors. 

The uncertainties in these backgrounds can be reduced with better knowledge of the  form factors and  lineshapes.
At high $q^2$ contributions from the 1430-family of higher kaon resonances are in addition kinematically
suppressed. For non-resonant contributions at low $q^2$ presently no theoretical calculation is available.
The new observable $A_{\rm FB \, L}^K$ (\ref{eq:AFB}) and (\ref{btilde})
shown in Figs.~\ref{pb2} and \ref{fig:btilde}, respectively, which arises from  interference between the $K^\ast$  and the background amplitudes, can be used to experimentally constrain
hadronic backgrounds efficiently irrespective of the underlying short-distance model.

Due to the narrower width and absence of prominent low mass $\bar s s$-scalars decaying to $KK$, the backgrounds in $B_s \to \Phi  (\to KK) \nu \bar \nu$ decays
are smaller than in $B^0 \to K^*(\to K \pi) \nu \bar \nu$ decays. The problem of finite width-related backgrounds may of course be avoided altogether with the 
decay $B \to K \nu \bar \nu$, which, however, allows to measure only  its differential decay rate proportional to  $f_+^2 (q^2)|C_L +C_R|^2$.

There clearly is feedback from $b \to s \nu \bar \nu$ to  $b\to s \ell^+ \ell^-$ transitions and back: these modes are related by
form factors and other hadronic input, as well as by Wilson coefficients in $SU(2)_L$-symmetric SM extensions. On the other hand,
the dineutrino modes are not polluted by electromagentic effects, and they do probe flavor physics in a complementary way, notably, third generation leptons are included, and may shed light on ongoing and future tests of lepton-universality.

\acknowledgments
The work of  GH and IN is  supported  in part by the  {\it  Bundesministerium f\"ur Bildung und Forschung} (BMBF).

\begin{appendix}
\section{Polarization vectors}\label{App:Polarization vectors}
Our conventions for the metric and Levi-Civita tensors are $g_{\mu\nu}={\rm diag}(1,-1,-1,-1)$ and $\epsilon_{0123}=1$, respectively. The choice of the polarization vectors of the neutrino pair in the $B$-rest frame is given by
  \begin{equation}
 \tilde{\epsilon}^\mu_{\pm}= \frac{1}{\sqrt{2}}(0,\mp 1, i,0),\,\,  \tilde{\epsilon}^\mu_0= \frac{1}{\sqrt{q^2}}(\vert \vec{q}\vert,0,0,q_0),\,\, \tilde{\epsilon}^\mu_{t}=\frac{1}{\sqrt{q^2}}(q^0,0,0,\vert \vec{q}\vert),
 \end{equation}
while the components of the four-vector $q^\mu=(q^0,0,0,\vert \vec{q}\vert)$, with $ \vert\vec{q}\vert=\sqrt{\lambda(m_B^2,q^2,m_{K^\ast}^2)}/(2m_B)$. 
The polarization vectors satisfy the conditions of orthogonality and completeness, respectively:
\begin{equation}
\begin{split}
& \tilde{\epsilon}^{\ast\,\mu}_m \tilde{\epsilon}_{\mu\,m'}= g_{m m'},\quad \sum_{m,m'} \tilde{\epsilon}^{\ast\,\mu}_m \tilde{\epsilon}^{\nu}_{m'}g_{m m'}=g^{\mu\nu}.
\end{split}
\end{equation} 
The direction of the $z$ axis is opposite to the direction of motion of the $K^\ast$ in the $B$-rest frame.
Polarization vectors of the $K^\ast$ meson in this frame read
 \begin{equation}
 \epsilon^\mu_\pm= \frac{1}{\sqrt{2}}(0,\pm 1, i,0),\,\, \epsilon^\mu(0)= \frac{1}{m_{K^\ast}}(k_z,0,0,k_0),
 \end{equation}
 where $k_z=-\vert \vec{q}\vert$. These satisfy the orthogonality and completeness relations
 \begin{equation}
 \begin{split}
&\epsilon^{\ast\,\mu}_m \epsilon_{\mu\, m'}=-\delta_{m m'},\quad \sum_{m, m'}\epsilon^{\ast\,\mu}_m \epsilon^\nu_{m'}\delta_{m m'}=-g^{\mu\nu}+\frac{k^\mu k^\nu}{m_{K^\ast}^2}.\label{completness-Kst}
 \end{split}
 \end{equation}
where the indices $m=0,1,2,3$ are ordered as $m=t; 0, \pm$, respectively.

\section{Form factors}\label{App:B}
The matrix elements of the vector and axial currents  $\bar{s}\gamma_\mu(\gamma_5)b$ between the $B$- and the  $K^\ast$-meson with polarization $n$ are parametrised with the standard form factors $V(q^2)$ and $A_{0,1,2}(q^2)$
\begin{equation}
\begin{split}
\langle K^\ast(k,n)\vert \bar{s}\gamma_\mu b\vert B(p_B)\rangle &=\epsilon_{\mu\nu\alpha\beta}\epsilon^{\ast\,\nu}_n p_B^\alpha k^\beta \frac{2 i V(q^2)}{m_B+m_{K^\ast}},\\
\langle K^\ast(k,n)\vert \bar{s}\gamma_\mu\gamma_5 b\vert B(p_B)\rangle&=-\,\epsilon^\ast_{n\,\mu}(m_B+m_{K^\ast})A_1(q^2)+(p_{B\,\mu}+k_\mu)\frac{\epsilon^\ast_n\cdot q}{m_B+m_{K^\ast}}A_2(q^2)+\\
&+q_\mu (\epsilon^\ast_n\cdot q) \frac{2m_{K^\ast}}{q^2}(A_3(q^2)-A_0(q^2)),
\end{split}
\end{equation}
where 
\begin{equation}
A_3(q^2)=\frac{m_B+m_{K^\ast}}{2m_{K^\ast}}A_1(q^2)-\frac{m_B-m_{K^\ast}}{2m_{K^\ast}}A_2(q^2).
\end{equation}
The matrix element for the final state scalar reads
\begin{equation}
\langle K_0^\ast(k)\vert \bar{s}\gamma_\mu \gamma_5 b\vert B(p_B)\rangle=\,(p_{B}+k)_\mu  f_+(q^2)+\,q_\mu f_-(q^2).\label{scalar-form-factor}
\end{equation}
The form factors for $B\to K^\ast_0(1430)$ are calculated within
QCD sum rules (QCDSR)~\cite{Aliev:2007rq}.  
In our numerical analysis we use QCDSR from factors which are parameterized  as~\cite{Aliev:2007rq}
\begin{equation}
 f_i({q}^2) = \frac{f_i(0)}{1-a_i ({q}^2/m_B^2) + b_i ({q}^2/m_B^2 )^2 } \, ,\label{scalar-ff}
\end{equation}
where $a_i, b_i$ for $i = +,-$ are fit coefficients. In the limit of vanishing lepton masses only  $f_+({q}^2)$ contributes. The corresponding parameters are $f_+(0) = 0.31\pm0.08$, $a_+=0.81$, $b_+=-0.21$~\cite{Aliev:2007rq}.

\section{Numerical input}\label{App:C}
In the following table we collect the numerical values of the inputs used in this paper.
\begin{table}[H]
\begin{center}
\begin{tabular}{c|c|c}
\hline\hline
Parameter           &     Value & Source \\\hline\hline
$|V_{ts}^* V_{tb}|$ & $0.0401 \pm 0.0010$ &  \cite{Bona:2006ah}\\
$\alpha_e(m_b)$ & 1/127.925(16) & \cite{PDG} \\
$\Gamma(B_0)$ & $(4.333 \pm 0.020) \cdot 10^{-13}$ GeV & \cite{PDG} \\
$\Gamma(B_s)$ & $(4.342 \pm 0.032) \cdot 10^{-13}$ GeV  & \cite{PDG} \\
$m_{K_0^\ast}$ & $1425\pm 50\,\text{MeV}$  & \cite{PDG}\\
$\Gamma_{K_0^\ast}$ & $270\pm 80\,\text{MeV}$  & \cite{PDG}\\
$m_\kappa$ & $658(13)\,\text{MeV}$ & \cite{DescotesGenon:2006uk}\\
$\Gamma_\kappa$ & $557(24)\,\text{MeV}$ & \cite{DescotesGenon:2006uk}\\
$\vert g_\kappa\vert$ & $[0  \ldots 0.2]$ & \cite{Becirevic:2012dp}\\
$\arg g_\kappa$ & $[\pi/2 \ldots  \pi]$ & \cite{Becirevic:2012dp}\\
$f_\pi $ & $130.4 \pm 0.2$ MeV & \cite{PDG} \\
$f_K $ & $156.2 \pm 0.7$ MeV &  \cite{PDG}$^\dagger$ \\
$f_{B_d} $ & $188 \pm 4$ MeV &  \cite{Dowdall:2013tga}\\
 $f_{B_s} $ & $224\pm 5$ MeV &  \cite{Dowdall:2013tga}\\
$g$ & $0.569 \pm 0.076$ & \cite{Flynn:2013kwa, Flynn:2015xna}$^\dagger$
\\\hline\hline
\end{tabular}
\end{center}
\caption{\label{tab:input}
$^\dagger$Uncertainties added in quadrature.}
\end{table}

\end{appendix}

\end{document}